\documentclass[reprint,twocolumn,prl,showpacs,amsmath,amssymb,superscriptaddress,nofootinbib,longbibliography]{revtex4-2}

\usepackage{graphicx}% Include figure files
\usepackage{dcolumn}% Align table columns on decimal point
\usepackage{bm}% bold math
\usepackage{booktabs}
\usepackage{placeins}
\usepackage{rotating}
\begin{document}

%\preprint{To be submitted to Physical Review Letters}
%\preprint{APS/123-QED}

%\title{Double-magicity of $^{52}$Ca resolved by precision mass measurements of neutron-rich scandium isotopes around $N=32$}% Force line breaks with \\
\title{Precision mass measurements of neutron-rich scandium isotopes refine the evolution of $N=32$ and $N=34$ shell closures}% Force line breaks with \\

\author{E. Leistenschneider}
\email[Corresponding author: ]{leistenschneider@frib.msu.edu}
    \affiliation{Facility for Rare Isotope Beams, Michigan State University, East Lansing, Michigan 48824, USA}
    \affiliation{National Superconducting Cyclotron Laboratory, Michigan State University, East Lansing, Michigan 48824, USA}

 \author{E. Dunling}
    \affiliation{TRIUMF, 4004 Wesbrook Mall, Vancouver, British Columbia V6T 2A3, Canada}
    \affiliation{Department of Physics, University of York, York, YO10 5DD, United Kingdom}

\author{G. Bollen}
    \affiliation{Facility for Rare Isotope Beams, Michigan State University, East Lansing, Michigan 48824, USA}
    \affiliation{National Superconducting Cyclotron Laboratory, Michigan State University, East Lansing, Michigan 48824, USA}
    \affiliation{Department of Physics and Astronomy, Michigan State University, East Lansing, Michigan 48824, USA}

\author{B.A. Brown}
    \affiliation{Facility for Rare Isotope Beams, Michigan State University, East Lansing, Michigan 48824, USA}
    \affiliation{National Superconducting Cyclotron Laboratory, Michigan State University, East Lansing, Michigan 48824, USA}
    \affiliation{Department of Physics and Astronomy, Michigan State University, East Lansing, Michigan 48824, USA}

\author{J. Dilling}
    \affiliation{TRIUMF, 4004 Wesbrook Mall, Vancouver, British Columbia V6T 2A3, Canada}
    \affiliation{Department of Physics \& Astronomy, University of British Columbia, Vancouver, British Columbia V6T 1Z1, Canada}

\author{A. Hamaker}
    \affiliation{Facility for Rare Isotope Beams, Michigan State University, East Lansing, Michigan 48824, USA}
    \affiliation{National Superconducting Cyclotron Laboratory, Michigan State University, East Lansing, Michigan 48824, USA}
    \affiliation{Department of Physics and Astronomy, Michigan State University, East Lansing, Michigan 48824, USA}
    
\author{J.D.~Holt}
    \affiliation{TRIUMF, 4004 Wesbrook Mall, Vancouver, British Columbia V6T 2A3, Canada}    
    \affiliation{Department of Physics, McGill University, 3600 Rue University, Montr\'eal, QC H3A 2T8, Canada}
      
\author{A. Jacobs}
    \affiliation{TRIUMF, 4004 Wesbrook Mall, Vancouver, British Columbia V6T 2A3, Canada}
    \affiliation{Department of Physics \& Astronomy, University of British Columbia, Vancouver, British Columbia V6T 1Z1, Canada}

\author{A.A. Kwiatkowski}
    \affiliation{TRIUMF, 4004 Wesbrook Mall, Vancouver, British Columbia V6T 2A3, Canada}
    \affiliation{Department of Physics and Astronomy, University of Victoria, Victoria, British Columbia V8P 5C2, Canada}      
    
\author{T. Miyagi}
    \affiliation{TRIUMF, 4004 Wesbrook Mall, Vancouver, British Columbia V6T 2A3, Canada}    
     %Takayuki Miyagi tmiyagi@triumf.ca   
    
\author{W.S. Porter}
    \affiliation{TRIUMF, 4004 Wesbrook Mall, Vancouver, British Columbia V6T 2A3, Canada}
    \affiliation{Department of Physics \& Astronomy, University of British Columbia, Vancouver, British Columbia V6T 1Z1, Canada}    

\author{D. Puentes}
    \affiliation{Facility for Rare Isotope Beams, Michigan State University, East Lansing, Michigan 48824, USA}
    \affiliation{National Superconducting Cyclotron Laboratory, Michigan State University, East Lansing, Michigan 48824, USA}
    \affiliation{Department of Physics and Astronomy, Michigan State University, East Lansing, Michigan 48824, USA}

\author{M. Redshaw}
    \affiliation{Department of Physics, Central Michigan University, Mount Pleasant, Michigan 48859, USA}
    \affiliation{National Superconducting Cyclotron Laboratory, Michigan State University, East Lansing, Michigan 48824, USA}    

\author{M.P. Reiter}
    \affiliation{TRIUMF, 4004 Wesbrook Mall, Vancouver, British Columbia V6T 2A3, Canada}
    \affiliation{II. Physikalisches Institut, Justus-Liebig-Universit\"{a}t, 35392 Gie{\ss}en, Germany}
    \affiliation{School of Physics and Astronomy, University of Edinburgh, Edinburgh, EH9 3FD, United Kingdom}

\author{R. Ringle}
    \affiliation{Facility for Rare Isotope Beams, Michigan State University, East Lansing, Michigan 48824, USA}
    \affiliation{National Superconducting Cyclotron Laboratory, Michigan State University, East Lansing, Michigan 48824, USA}

\author{R. Sandler}
    \affiliation{Department of Physics, Central Michigan University, Mount Pleasant, Michigan 48859, USA}

\author{C.S. Sumithrarachchi}
    \affiliation{Facility for Rare Isotope Beams, Michigan State University, East Lansing, Michigan 48824, USA}
    \affiliation{National Superconducting Cyclotron Laboratory, Michigan State University, East Lansing, Michigan 48824, USA}

\author{A.A. Valverde}
    \affiliation{Department of Physics \& Astronomy, University of Manitoba, Winnipeg, Manitoba R3T 2N2, Canada}
    %\affiliation{Physics Division, Argonne National Laboratory, Lemont, Illinois 60439, USA}

\author{I.T. Yandow}
    \affiliation{Facility for Rare Isotope Beams, Michigan State University, East Lansing, Michigan 48824, USA}
    \affiliation{National Superconducting Cyclotron Laboratory, Michigan State University, East Lansing, Michigan 48824, USA}
    \affiliation{Department of Physics and Astronomy, Michigan State University, East Lansing, Michigan 48824, USA}

\author{the TITAN Collaboration} \noaffiliation

%\collaboration{TITAN Collaboration}%\noaffiliation

%\affiliation{
%\affiliation{
%\affiliation{%

\date{\today}% It is always \today, today,

\begin{abstract}
%Emerging closed shell effects among nuclei with 32 neutrons has been object of great curiosity.  
%.............
We report high-precision mass measurements of $^{50-55}$Sc isotopes performed at the LEBIT facility at NSCL and at the TITAN facility at TRIUMF. Our results provide a substantial reduction of their uncertainties and indicate significant deviations, up to 0.7 MeV, from the previously recommended mass values for $^{53-55}$Sc. The results of this work provide an important update to the description of  emerging closed-shell phenomena at neutron numbers $N=32$ and $N=34$ above proton-magic $Z=20$.
In particular, they finally enable a complete and precise characterization of the trends in ground state binding energies along the $N=32$ isotone, confirming that the empirical neutron shell gap energies peak at the doubly-magic $^{52}$Ca. Moreover, our data, combined with other recent measurements, does not support the existence of closed neutron shell in $^{55}$Sc at $N=34$. The results were compared to predictions from both \emph{ab initio} and phenomenological nuclear theories, which all had success describing $N=32$ neutron shell gap energies but were highly disparate in the description of the $N=34$ isotone.   %Further, we also discuss the implications of our results for the existence of any shell effect at $N=34$.
 %Further, we discuss potential implications for the theoretical description of the neighboring emerging shell, at $N=34$.

%\begin{description}
%\pacs{21.10.Dr}
%\end{description}
\end{abstract}

%\pacs{21.10.Dr}% PACS, the Physics and Astronomy
\maketitle

The formation of simple, periodic patterns is a key to understanding the organization of countless many-body systems \cite{Ruiz2020}. In the case of atomic nuclei, clear repetition of special properties, such as enhanced binding energy, are seen in ``magic'' proton and neutron numbers (like 2, 8, 20, 28, 50...), which led to the proposition that nucleons organize themselves into shell-like structures, analogous to atomic electron orbitals. Today, the Nuclear Shell Model \cite{Mayer-Jensen} constitutes the foundation of our understanding of these objects. However, once believed to be immutable, these magic nucleon numbers are now known to vanish and new ones to appear in extreme cases of proton-to-neutron ratio \cite{Warner2004}. %, which  give us insights on how nucleons interact to each other as the nuclear environment changes. 
The appearance and evolution of emerging shell closures have become standard metrics to benchmark nuclear theories \cite{Otsuka2020,Wienholtz2013}.

Emerging closed shell phenomena have been observed for neutron-rich nuclei at neutron numbers $N=32$ and $N=34$ at and around the proton-magic calcium chain ($Z=20$). Such behaviors at $N=32$ were found through several observables. Mass spectrometry experiments identified peaks in empirical neutron shell gap energies in $^{51}$K ($Z=19$) \cite{Rosenbusch2015}, $^{52}$Ca ($Z=20$) \cite{Gallant2012a,Wienholtz2013}, $^{53}$Sc ($Z=21$) \cite{Xu2019}, and $^{54}$Ti ($Z=22$) \cite{Leistenschneider2018}, but not in  $^{55}$V ($Z=23$) \cite{Reiter2018} and higher proton numbers. This shell closure is also seen in enhanced $2^+$ excitation energies at $^{50}$Ar ($Z=18$) \cite{Steppenbeck2015}, $^{52}$Ca \cite{Huck1985}, $^{54}$Ti \cite{Liddick2004}, and  $^{56}$Cr ($Z=24$) \cite{PRISCIANDARO2001}, and reduced $B(E2)$ transition probabilities in $^{54}$Ti \cite{Goldkuhle2019} and $^{56}$Cr \cite{Seidlitz2011}. In $N=34$, albeit less experimentally studied, evidences of a shell closure were observed in the mass of $^{54}$Ca \cite{Michimasa2018} and in the $2^+$ excitation energies of $^{54}$Ca \cite{Steppenbeck2013,Chen2019} and $^{52}$Ar \cite{Liu2019}. %although . but the scarcity of experimental data puts into question whether it exists above $Z=20$ \cite{Steppenbeck2017}.

In the Shell Model, a magic number appears when there is a significant energy gap between two nucleon orbitals. In $N=32$ nuclei with more than 24 protons, neutrons in the valence $p_{3/2}$ orbital are quasi-degenerate with those in the next orbital -- $f_{5/2}$ -- thus no appreciable energy gap is observed. As protons are removed, a strong residual interaction between protons and neutrons weakens, causing a migration of the neutron $f_{5/2}$ energy level, a widened gap with the $p_{3/2}$, and thus the formation of the $N=32$ shell closure \cite{Otsuka2020}. Subsequently, as proton number drops below $Z=23$ and the proton-neutron residual interaction further wanes, an energy level inversion between the neutron orbitals $f_{5/2}$ and $p_{1/2}$ occurs \cite{Liddick2004, Steppenbeck2013}. The next neutron orbital -- $p_{1/2}$ -- has then degeneracy of 2, allowing also for the formation of the $N=34$ shell closure provided that a significant gap can be formed with the $f_{5/2}$ orbital.

%The current Shell model picture intimately connects both emergent phenomena at $N=32$ and $34$ \cite{Otsuka2020}. Above $Z=24$, neutrons in the valence $f_{5/2}$ orbital are quasi-degenerate with those in the $p_{3/2}$ and strongly interact with the protons in the valence $f_{7/2}$ orbital. As protons are removed, this interaction weakens, causing a migration of the neutron $f_{5/2}$ energy level and a widened gap with the $p_{3/2}$, thus forming  characteristic closed shell features at $N=32$. Subsequently, as proton number drops below $Z=23$ \cite{Liddick2004, Steppenbeck2013} and the proton-neutron residual interaction further wanes, an energy level inversion between the neutron orbitals $f_{5/2}$ and $p_{1/2}$ allows the formation of the $N=34$ shell. 

The conjoint structural changes at $N=32$ and $34$ provide a unique opportunity to refine our understanding of nuclear interactions and the intricate nuclear many-body problem \cite{Otsuka2020,Wienholtz2013, Leistenschneider2018}. Therefore, detailed information on how the characteristic observables evolve with increments in proton number is of paramount importance. Above $Z=20$, the evolution of the $N=32$ shell closure is nearly resolved, %among energy observables, such as masses and excitation energies, 
but a few outstanding issues remain.  Among them, recent Isochronous Mass Spectrometry (IMS) measurements suggest the empirical neutron shell gap at $N=32$ is unexpectedly higher in $^{53}$Sc than in the doubly-magic $^{52}$Ca  \cite{Xu2015,Xu2019}. In $N=34$, experimental evidences indicate an absence of closed shell signatures in Ti  \cite{Liddick2004,Dinca2005} and their presence in Ca \cite{Steppenbeck2013,Michimasa2018,Chen2019} but, in between, the picture in Sc remains unclear \cite{Crawford2010,Steppenbeck2017,Meisel2020}.

These questions demand refined mass measurements of neutron-rich scandium isotopes, which remains the only chain in the region still unexplored using high-resolution techniques. 
In this letter, we report precision mass measurements of $^{50-55}$Sc, between $N=29$ and $34$, performed in a joint collaboration between experimental groups at the National Superconducting Cyclotron Laboratory (NSCL) in the U.S. and at the TRIUMF National Laboratory in Canada. %With our results, the mass surface around $N=32$ reveals a smooth evolution of the shell gaps at the $N=32$ isotone, and confirms that empirical shell gaps indeed peak at the doubly-magic $^{52}$Ca. 

%Our data finally enable complete mass characterisation of closed shell behaviors in $N=32$ above $Z=20$. % the completes the study of  last isotopic chain still unexplored with high-resolution techniques for 

At NSCL, the neutron-rich isotopes $^{50-53}$Sc were produced in flight by nuclear fragmentation of a 130 MeV/u $^{76}$Ge primary ion beam impinging on a natural Be target of about  0.4 g/cm$^2$ thickness. The beam was purified at the A1900 fragment separator \cite{MORRISSEY2003} and delivered to the NSCL's gas catcher \cite{SUMITHRARACHCHI2020}, where the high-energy fragments were stopped in a high-purity He gas. The ions were extracted at low  energies from the gas cell and selected in mass-to-charge ratio ($A/Q$) by a dipole magnet. The species of interest were extracted as singly charged molecules, mostly oxides, formed during stopping in the gas cell. The ion beam was then delivered to the Low Energy Beam and Ion Trap (LEBIT) facility \cite{RINGLE2013}.

LEBIT is an ion trap facility dedicated to performing high-precision mass spectrometry of short-lived ions \cite{RINGLE2013}. The beam was received into LEBIT's cooler and buncher \cite{SCHWARZ2016}, where the continuous rare isotope beam was converted into short low-emittance bunches. The ion bunches were then sent to LEBIT's 9.4 T Penning trap mass spectrometer \cite{RINGLE2009},  where they were further purified against isobaric contaminants by applying a dipolar radio-frequency (RF) field \cite{Kretzschmar2013,KWIATKOWSKI2015}.

\begin{figure}[]
    \begin{center}
        \includegraphics[width=\columnwidth]{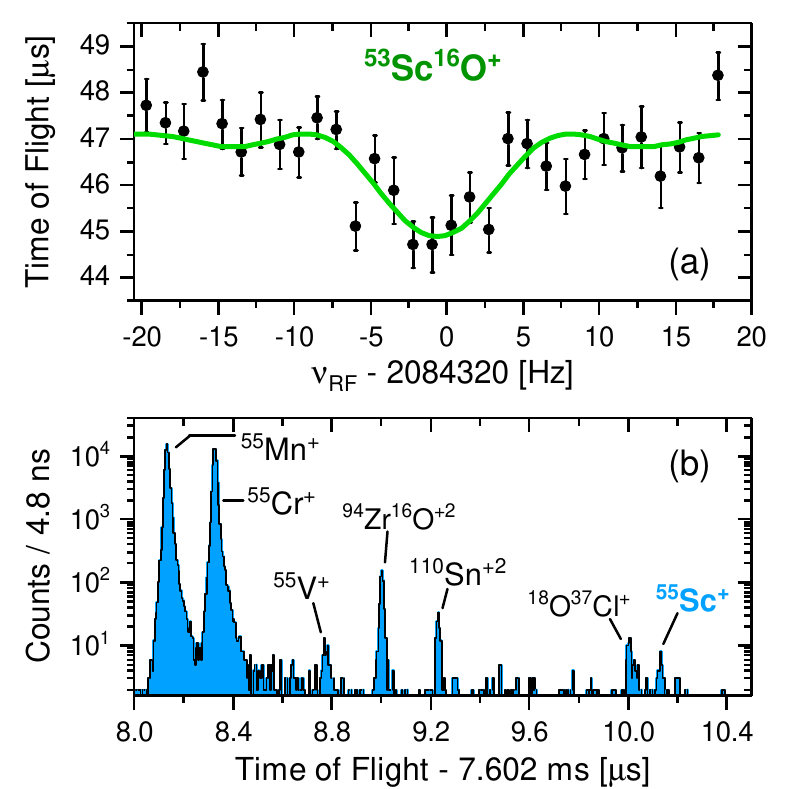}
        \caption{A sample mass spectrum from each experiment: (a) A ToF-ICR resonance of a $^{53}$Sc$^{16}$O$^{+}$ molecular ion, obtained with excitation time of 100 ms at LEBIT. The dip in time of flight occurs at $\nu_c$, extracted using an analytical fit (green).  (b) A typical MR-TOF-MS spectrum at $A/Q=55$ obtained at TITAN with 520 isochronous turns. }
        \label{fig:spectra}
        %\vspace*{-8mm}
    \end{center}
\end{figure}

The measurement of the mass ($m_{ion}$) of the ion is done through the measurement of the frequency ($\nu_c$) of the cyclotron motion about the trap's magnetic field: $\nu_c  = {(q  \, B)}/{(2\pi \, m_{ion})} $, where $q $ is the charge of the ion and $B$ is the strength of the magnetic field.  To measure $\nu_c$, we employed the Time-of-Flight Ion-Cyclotron-Resonance technique (ToF-ICR) \cite{Konig1995}, using standard quadrupole excitation schemes ranging between 50 ms and 500 ms, depending on the measurement. Figure \ref{fig:spectra}.a shows a typical ToF-ICR spectrum obtained for $^{53}$Sc$^{16}$O$^{+}$.

The calibration of $B$ was done through the measurement of the cyclotron frequency of a reference ion ($\nu_{c,\text{ref}}$), whose mass has been precisely measured and is well documented in the literature \cite{AME16}.  The reference ions were all well-known molecular ions produced in the gas catcher and delivered with the ion of interest, with the same $A/Q$. %, to avoid mass-dependant systematic deviations.  
Measurements of $\nu_{c,\text{ref}}$ were performed at intervals not longer than 1.5 hour, interleaved between measurements of $\nu_c$ of ions of interest, to account for temporal variations in the magnetic field.  

\begin{table*}[ht]
  \centering
  \caption{Results of the mass measurements performed, compared to the values recommended by the AME16 \cite{AME16}. We also provide the mass ratio between the ion of interest and the reference ion, which is equivalent to the average cyclotron frequency ratio (R) for the LEBIT ToF-ICR data. All mass values are in keV. }
    \begin{tabular}{l c c c c c c c}
    \toprule    Facility & Nuclide & Mass Excess   & Literature \cite{AME16} & Difference & $\quad$ Ion of Interest   & Reference Ion & Mass Ratio \\    \hline
LEBIT & $^{50}$Sc &  -44537.1 (2.5) &  -44547 \, (15) & 10 (15) &  	$\quad$ $^{50}$Sc\,$^{16}$O\,$^{+}$		 &  $^{12}$C\,$^{19}$F\,$^{35}$Cl\,$^{+}$	 	&  0.999\,694\,485 (41)	  \\

 & $^{51}$Sc 	&  -43250.4 (2.5) & -43229 \, (20) & -22 (20) &  	$\quad$ $^{51}$Sc\,$^{16}$O\,$^{1}$H\,$^{+}$	 &  $^{33}$S\,$^{19}$F\,$^{16}$O\,$^{+}$	&  0.999\,875\,402 (39)  \\
 
 & $^{52}$Sc &  -40523.6 (3.0) & -40443 \, (82) & -80 (82) &  $\quad$ 	$^{52}$Sc\,$^{16}$O\,$^{+}$		 &  $^{33}$S\,$^{19}$F\,$^{16}$O\,$^{+}$	&  0.999\,803\,339 (48)	 \\
 
 & $^{53}$Sc &  -38769 \,\,\, (17)  & -38907  \, (94) & 138 (96)  &  	$\quad$ $^{53}$Sc\,$^{16}$O\,$^{+}$		 &  $^{34}$S\,$^{19}$F\,$^{16}$O\,$^{+}$	&  0.999\,885\,58\,\,\,  (27)	\\ \hline

 TITAN & $^{54}$Sc &  -34438 \,\,\, (18)  &  -33891 (270) & -547 (270) &  $\quad$ 	$^{54}$Sc$^{+}$		 &  $^{54}$Cr$^{+}$	 	&  1.000\,447\,76\,\,\, (36)
	\\
 
      & $^{55}$Sc &  -30842 \,\,\, (62) & -30159 (450) & -683 (460)  &  $\quad$ 	$^{55}$Sc$^{+}$		 &  $^{55}$Cr$^{+}$	 	&  1.000\,474\,2\,\,\,\,\,\, (12)
	   \\  \hline

% TITAN & $^{54}$Sc &  -34438 \,\,\, (18)  &  -33891 (270) & -547 (270) &  $\quad$ 	$^{54}$Sc$^{+}$		 &  $^{54}$Cr$^{+}$	 	&  22497 (18)	\\
 
%      & $^{55}$Sc &  -30842 \,\,\, (62) & -30159 (450) & -683 (460)  &  $\quad$ 	$^{55}$Sc$^{+}$		 &  $^{55}$Cr$^{+}$	 	&  24268 (62)	   \\  \hline

    \end{tabular}%
  \label{tab:results-thisexps}%
\end{table*}

%The analysis of ToF-ICR spectra followed similar procedures as other experiments performed at LEBIT, for example as in  \cite{Kandegedara2017}.  
Each ToF-ICR spectrum was fitted with an analytical function described in \cite{Konig1995}, from which the cyclotron frequency was obtained. A count-rate class analysis \cite{Kellerbauer2003} was performed to account for frequency shifts due to ion-ion interactions. The atomic mass of the species of interest ($m$) was determined from the ratio of cyclotron frequencies ($R$) of the ion of interest and the reference ion: $ R  = {\nu_{c,\text{ref}}}/{\nu_{c}}  = {(m- m_e)}/{(m_\text{ref}-  m_e)} $,  where $m_e$ is the mass of the electron and $m_\text{ref}$ is the mass of the reference species (obtained from \cite{AME16}, summing the masses of all atoms of the molecule). This equation   is valid for singly ionized species and it disregards insignificant electron and molecular binding energies, on the order of a few eV or less.  $\nu_{c,\text{ref}}$ was obtained from an interpolation, to the time of the measurement of the ion of interest, between the reference measurements before and after it.  The  mass of the nuclide of interest is obtained by subtracting the masses of its molecular counterparts.

%The masses of  $^{54}$Sc and  $^{55}$Sc were measured at TRIUMF's Ion Trap for Atomic and Nuclear science (TITAN) facility \cite{DILLING2006198}. %These measurements were part of an experimental campaign that aimed to extend past successful results in the same region \cite{Leistenschneider2018,Reiter2018}, and therefore follow very similar procedures. Further results of this campaign will be reported in a future publication.
At TRIUMF, samples of  $^{54}$Sc and  $^{55}$Sc were produced via the ISOL method through spallation reactions at the Isotope Separator and ACcelerator (ISAC) \cite{Ball2016} facility by impinging a 480 MeV proton beam of 50 $\mu$A onto a Ta target, 22.7 g/cm$^2$ thick. The reaction products were stopped and thermalized in the target material, released through desorption, and surface ionized at the TRILIS ion source \cite{Lassen2009}. The beam was extracted from the target, selected in $A/Q$ at ISAC's dipole mass separator \cite{Bricault2002} and delivered at low energy to TRIUMF's Ion Trap for Atomic and Nuclear science (TITAN) facility \cite{DILLING2006198}.

%TITAN, similarly to LEBIT, is an ion-trap facility dedicated to perform mass spectrometry of short-lived ions \cite{DILLING2006198}. It employs two mass spectrometers, a Penning Trap  Mass Spectrometer \cite{BRODEUR201220} and a Multiple-Reflection Time-of-Flight Mass Spectrometer (MR-TOF-MS) \cite{Jesch2015}. In this experiment, the radioactive beam from ISAC was accumulated at TITAN's cooler and buncher \cite{BRUNNER201232} for 20 ms and sent as ion bunches to the MR-TOF-MS. 

The radioactive beam was accumulated at TITAN's cooler and buncher \cite{BRUNNER201232} for 20 ms and sent as ion bunches to the Multiple-Reflection Time-of-Flight Mass Spectrometer (MR-TOF-MS) \cite{Jesch2015}. The MR-TOF-MS determines the mass of a charged particle through its time-of-flight through a standardized path at known kinetic energy  \cite{WOLLNIK1990,PLASS2013134}. In order to increase the measurement resolution, this device confines the ion by bouncing it between a pair of electrostatic mirrors, recycling and thus extending the flight path and preserving the initial time spread. In this experiment, the bunches were received by the MR-TOF-MS in an internal ion preparation system composed by gas-filled RF quadrupoles \cite{Jesch2015}, where ions were re-cooled for about 13 ms.  The ions were then sent to the mass analyser for times of flight of about 7.5 ms, where they underwent 520 isochronous turns between the mirrors. Finally, they were detected by a MagneTOF detector \cite{Dickel2019}.  An in-analyser mass-range selector, similar to \cite{DICKEL2015}, was used to remove any particle outside the desired $A/Q$ window. Also, to prevent ion-ion interaction effects, the average count rate was kept below 2 counts/cycle.  

A typical time of flight spectrum is shown in figure \ref{fig:spectra}.b. Every peak in the MR-TOF-MS spectra was fitted using a Gaussian function. The spectra were mass calibrated using the nonrelativistic relationship: ${m_{ion}}/{q} = C \, (t_{tof}-t_0)^2$, where $C$ and $t_0$ are calibration constants and $t_{tof}$ is the fitted centroid of the peak. The parameter $t_0$ is a constant delay caused in the signal processing and was determined using offline measurements. In both spectra taken with $A/Q = 54$ and $55$, we identified the presence of isobaric singly charged Cr, Fe, Mn, V, Ti, Sc, and several molecules.  Stable $^{54,55}$Cr$^+$ formed dominant peaks in their spectra and, therefore, were chosen as suitable calibrants for $C$ using their atomic mass values available in the literature \cite{AME16}. A time-dependent calibration, similar to \cite{Ayet2019}, was used to account for drifts in time of flight. A mass resolving power of about $200 \, 000$ was achieved, and a total of 236 counts of $^{54}$Sc and 35 counts of $^{55}$Sc were registered.

The atomic masses of the species of interest were determined through the relationship $m = m_{ion}+m_e$, which disregards electron binding energies. A relative systematic uncertainty of $3 \cdot 10^{-7}$ was added following the prescription outlined in previous experiments \cite{Leistenschneider2018,Reiter2018,Reiter2020}, as well as an additional relative  uncertainty of $1.9 \cdot 10^{-7}$ to account for ion-ion interaction effects   \cite{Dunling2020}. 

The masses obtained in both experiments are reported in table \ref{tab:results-thisexps} with comparisons with the recommended values by the Atomic Mass Evaluation of 2016 (AME16) \cite{AME16}. The precision was improved in all cases, some by over an order of magnitude, to the scale of a few tens of keV or better. %Our measurements are compatible, within 2.0 $\sigma$, with previous experimental results employing moderate-resolution techniques, including the two most recent  \cite{Xu2019,Meisel2020}. These two are re-analysis of data acquired in previous experiments \cite{Xu2015,Meisel2015}, which dominate the recommended values by the Atomic Mass Evaluation of 2016 (AME16) \cite{AME16}. Despite the 2.0 $\sigma$ compatibility with previous experiments, 
Deviations were found, up to 0.7 MeV, in the masses of $^ {53-55}$Sc ($N=32-34$), which can significantly impact the description of nuclear structure phenomena.

\begin{figure}[]
    \begin{center}
        \includegraphics[width=0.93\columnwidth]{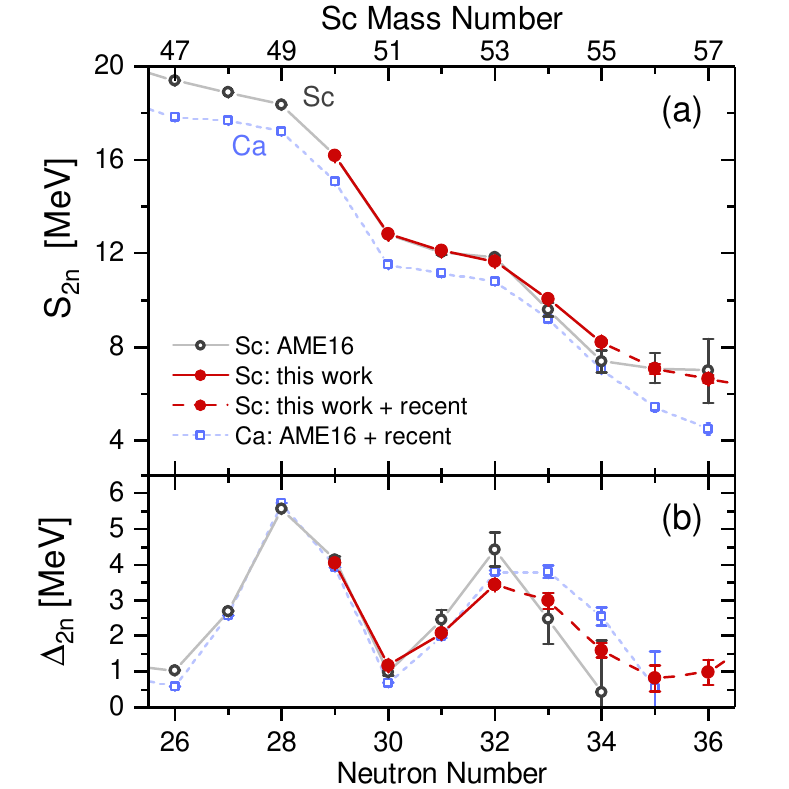}
        \caption{Derivatives of the mass surface around neutron numbers $N=28,32,34$:  (a) $S_{2n}$ and (b) $\Delta_{2n}$ of Sc isotopes as a function of $N$. Our results (red circles) provide a $2\sigma$ lower $\Delta_{2n}$ at $N=32$ compared to the AME16 \cite{AME16} (gray). Results from \cite{Meisel2020,Michimasa2020} were used to compound our $\Delta_{2n}$ at $N\geqslant33$ (dashed).  The proton-magic calcium chain (blue squares) was added for reference, with data from \cite{AME16,Michimasa2018}.   }
        \label{fig:masssurface}
    \end{center}
\end{figure}

%Signatures of shell effects can be identified as sharp features in the mass surface. 
Due to the strong binding character of closed shells, nucleons added on top of them are appreciably less bound.  
Hence, closed neutron shells can be identified by sudden decreases in two-neutron separation energies, defined as: $ S_{2n} (Z,N) =  \left[ m(Z,N-2) + 2m_n  - m(Z,N) \right] c^2 $, where $m_n$ is the mass of the neutron. These changes in slopes are emphasized by looking at derivatives of $S_{2n}$,  known as empirical shell gaps: $ \Delta_{2n} (Z,N) =  S_{2n}(Z,N)-  S_{2n}(Z,N+2)  $. Figure \ref{fig:masssurface} shows $S_{2n}$ and $\Delta_{2n}$ for scandium isotopes employing our values and AME16 values \cite{AME16}. Consecutive sharp features are clearly seen at $N=28$, a ``canonical'' magic number, and at $N=32$, an ``emerging'' magic number.  Our results confirm a strong $N=32$ subshell closure in $^{53}$Sc, but provide a $1$ MeV lower (2$\sigma$ away) empirical neutron shell gap energy ($3.45\pm 0.06$ MeV) than the AME16 and the most recent experimental result \cite{Xu2019} ($4.4 \pm 0.5$ MeV). 

\begin{figure*}[]
    \begin{center}
        \includegraphics[width=17.5cm]{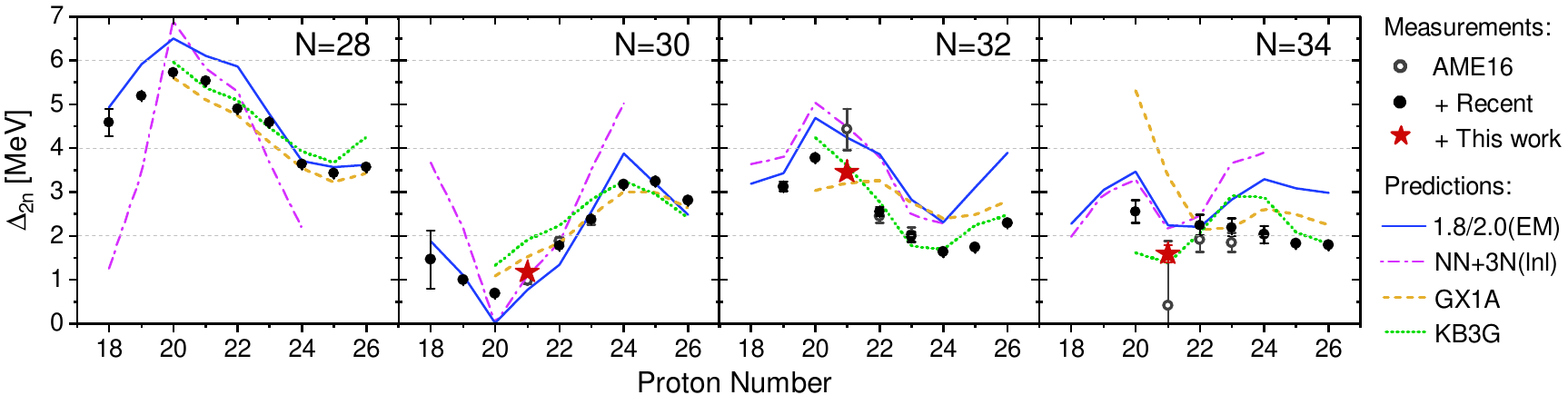}
        \caption{Measured (points) and predicted (lines) empirical neutron shell gaps for isotones in the region of interest: the canonical magic $N=28$, the non-magic $N=30$, the emerging magic $N=32$, and the possibly magic $N=34$. Measured points include the AME16 \cite{AME16} (open circles), with recent measurements \cite{Leistenschneider2018,Reiter2018,Michimasa2018,Michimasa2020} (full circles), and our work (with \cite{Meisel2020,Michimasa2020}) (red stars).  Predictions include VS-IMSRG calculations using 1.8/2.0(EM) (solid blue) and  NN+3N(lnl) (pink dash-dots) chiral Hamiltonians, and shell model calculations using KB3G (green dots) and  GX1A (yellow dashes) phenomenological Hamiltonians. }
        \label{fig:shelgapstheory}
    \end{center}
\end{figure*}

Our update to the empirical neutron shell gap of $^{53}$Sc considerably impacts the evolution of the $N=32$ shell closure and finally permits its inspection using only high-resolution mass data. Figure \ref{fig:shelgapstheory} shows $\Delta_{2n}$ as a function of $Z$ for  isotones of interest in the region, combining mass values from AME16 \cite{AME16}, recent experiments \cite{Leistenschneider2018,Reiter2018,Michimasa2018,Xu2019,Meisel2020} and our data. %the canonical magic $N=28$, the emerging magic $N=32$, and the non-magic $N=30$ isotones. 
Clearly, the shell gap at $N=32$ evolves smoothly without abrupt changes, mirroring the evolution of the canonical $N=28$ shell but $\approx2$ MeV lower. With our data, $\Delta_{2n}$ increases monotonically through $^{53}$Sc and peaks at $^{52}$Ca, consistent with the latter being a doubly-magic nucleus.

To derive $\Delta_{2n}$ towards $N=34$, we incorporated mass values of $^{56,57}$Sc from two recent Time-of-flight Magnetic Rigidity experiments at NSCL \cite{Meisel2020} and RIKEN \cite{Michimasa2020}. The trends in $S_{2n}$  in Sc seems to indicate  the restoration of non-closed shell slope after $N=34$. It contrasts with the Ca chain, where the continuation of a sharp slope through $N=34$ is an evidence of existence of a shell closure \cite{Michimasa2018}. We update the $\Delta_{2n}$ at $^{55}$Sc to $1.59 \pm 0.20$ MeV. This is the lowest empirical neutron shell gap in the $N=34$ isotone, as can be seen in Fig. \ref{fig:shelgapstheory}. High-resolution mass measurements of $^{56,57}$Sc are required to confirm the observed trends.   %     This result potentially indicate the presence of shell .....Although the uncertainty in the mass of $^{57}$Sc

As our data enables a more consistent picture of the evolution of both $N=32$ and $34$ isotones, we compared $\Delta_{2n}$ data with predictions from state-of-the-art nuclear models, which are included in figure \ref{fig:shelgapstheory}. We chose four approaches that had great success in describing closed shell features at $N=32$. Among phenomenological approaches, we performed shell model calculations using the KB3G \cite{POVES2001} and the  GX1A \cite{Honma2004} Hamiltonians for all nuclei  from Ca ($Z=20$) to Fe ($Z=26$) between $N=26$ and $N=36$, using the full $pf$ model space. Both interactions have been successful in predicting trends in binding energies \cite{Gallant2012a,Wienholtz2013, Michimasa2018,Meisel2020}, excitation energies \cite{Holt2012,Goldkuhle2019,Honma2005} and transition strengths \cite{Seidlitz2011,Goldkuhle2019} associated with emerging closed shells in the region. Among \emph{ab initio} approaches, we employed Valence-Space In-Medium Similarity Renormalization Group \cite{Stro16TNO,Stro17ENO} (VS-IMSRG) calculations with the 1.8/2.0(EM) \cite{Simo17SatFinNuc} and the NN+3N(lnl) \cite{Soma2020} interactions.  The 1.8/2.0(EM) interaction has been extensively employed in the region and adequately predicted the presence of the $N=32$ closed shell in the mass surface \cite{Simo17SatFinNuc,Leistenschneider2018,Mougeot2018,Xu2019,holt2019ab}. This approach has also had success in describing excitation energies around this isotone \cite{Simo17SatFinNuc,Steppenbeck2017,Liu2019}, and trends of charge radii in neutron-rich Ca isotopes \cite{GarciaRuiz2016}.  The NN+3N(lnl) interaction, released more recently, has also correctly predicted closed shell behaviors in the region \cite{Leistenschneider2018,Chen2019,Soma2020, Soma2020b} and has shown a comparable performance to the 1.8/2.0(EM) interaction. Our work is the first application of the NN+3N(lnl) interaction within the VS-IMSRG approach, which also enables the analysis of its performance across different many-body methods. 

Overall, the trends up to $N=32$ are well reproduced by all predictions. In particular, the KB3G interaction produces excellent  agreement with data, not only reproducing trends but also  magnitudes of shell gaps along most of the $N=28,32$ isotones.  Also, both VS-IMSRG calculations correctly reproduce the observed trends, although they overpredict the strength of shell gaps at $N=32$ \cite{Simo17SatFinNuc,Leistenschneider2018}. The calculations with the NN+3N(lnl) interaction have poorer performance at $N=28,30$ but produce nearly identical results as the 1.8/2.0(EM) interaction in the region of the emerging shell closures. The GX1A interaction, however, produces a less pronounced evolution of the $N=32$ isotone, peaking at $^{54}$Ti and not at the double-magic $^{52}$Ca. %In addition, it predicts a strong shell gap at $^{54}$Ca ($N=34$), which is also not observed.

%All theories predict the well-established double-magicity of $^{48}$Ca at $N=28$. The double-magicity of $^{52}$Ca at $N=32$ is also supported by VS-IMSRG and KB3G calculations. However, the GX1A interaction produces a less pronounced evolution of the $N=32$ isotone, peaking at $^{54}$Ti. In addition, it predicts a strong shell gap at $^{54}$Ca at $N=34$, which is also not observed.

%Emergent shell signatures at $N=34$ have also attracted attention in recent years \cite{Steppenbeck2013,Steppenbeck2017,Michimasa2018,Liu2019,Chen2019}. Shell model descriptions attribute the mechanisms that drive the appearance and evolution of shell effects in $N=32$ also to the $N=34$ \cite{Otsuka2020}. Therefore, it is natural to compare the performance of theoretical descriptions of these two isotones. 

The theoretical predictions are highly disparate regarding the description of the $N=34$ shell gap. % Although mass data are not refined enough to provide a clear description, shell gaps at $N=34$ do not indicate any meaningful shell effect at $Z\geq 20$. 
The GX1A interaction predicts a strong shell gap at $^{54}$Ca, which is not observed. The KB3G reproduces well $\Delta_{2n}$ in $^{55}$Sc and $^{56}$Ti, but does not predict the emergence of closed shell behaviors at $^{54}$Ca. A similar discrepancy has also been observed in descriptions of $2^+$ excitation energy using these interactions \cite{Honma2005,Holt2012,Steppenbeck2013}. The $2^+$ state of $^{54}$Ca lies $2.04\,(2)$ MeV  above the ground state \cite{Steppenbeck2013}. According to our calculations with the GX1A interaction, it lies at $2.96$ MeV, while KB3G predicts $1.34$ MeV.   Both VS-IMSRG calculations are in better agreement with $N=34$ data at $Z\leqslant 22$. Most remarkably, they reproduce the $\Delta_{2n}$ leap between $^{55}$Sc and $^{54}$Ca, but predict $\Delta_{2n}$  to reduce towards argon ($Z=18$). It  conflicts with recent $\gamma$-spectrometry results, that indicate  $N=34$ closed shell behaviors overcomes $N=32$ in strength at Ar  \cite{Steppenbeck2015,Liu2019}.

In summary, we performed high-precision mass measurements of $^{50-53}$Sc at the LEBIT facility at NSCL and of $^{54,55}$Sc at the TITAN facility at TRIUMF.  With our mass values, we obtained a smooth and monotonic evolution of the $N=32$ neutron shell gaps above $Z=20$, finally completing the mass description of the emergence of this closed shell using high-resolution methods. The observed behavior conforms to typical shell evolution as confirmed by various theoretical predictions, both from phenomenological and \emph{ab initio} approaches. Our results support $^{52}$Ca as a doubly-magic nucleus, establishing it as the peak of empirical shell gaps instead of $^{53}$Sc. Regarding the possibly emerging closed shell at $N=34$, our results combined with recent data from \cite{Meisel2020,Michimasa2020} suggest that closed-shell behaviors only appear in the mass surface at $Z \leqslant20$.   Our analysis also explored some theoretical approaches that have had superior performance in describing emerging closed shell behaviors in the region.  Despite the proposed intimate relationship between the emergence of closed shells in $N=32$ and $34$, success in describing observables in $N=32$ is not correlated of similar successes in $N=34$. %Naturally, we encourage  continuous exploration of this region using high-resolution mass spectrometry techniques, both towards the higher-$N$ and lower-$Z$ boundaries.

\begin{acknowledgments}
The authors would like to thank J. Simonis and P. Navr\'atil for providing the 1.8/2.0(EM) and NN+3N(lnl) matrix element files and S. R. Stroberg for the imsrg++ code \cite{Stro17imsrgplusplus} used to perform the VS-IMSRG calculations.  We are grateful to Z. Meisel for the fruitful discussions regarding data obtained in previous experiments. We also thank NSCL staff, the ISAC Beam Delivery group, the TRILIS group, and M. Good for their technical support, as well as J. Bergmann for his assistance with analysis software employed in this work.  This work was conducted with the support of Michigan State University, the U.S. National Science Foundation under Contracts No. PHY-1565546 and PHY-1811855,  the U.S. Department of Energy, Office of Science, Office of Nuclear Physics under Award No. DE-SC0015927,  the Natural Sciences and Engineering Research Council (NSERC) of Canada through Contract No SAPPJ-2018-00015 and the National Research Council (NRC) of Canada through TRIUMF. E.D. acknowledges financial support from the U.K.-Canada foundation. M.P.R. acknowledges support from BMBF (Grants No. 05P16RGFN1 and No. 05P19RGFN8), HMWK through the LOEWE Center HICforFAIR, by the JLU and GSI under the JLU-GSI strategic Helmholtz partnership agreement. A.A.V. acknowledges support from NSERC (Canada) under Contract No. SAPPJ-2018-00028.

\end{acknowledgments}

%\begin{thebibliography}
\bibliography{library}
%\end{thebibliography}

\end{document}